\documentclass[a4paper,11pt]{article}
\usepackage{pos}
\usepackage{siunitx}
\usepackage{tikz}

\title{Physics Performance and Detector Requirements at an Asymmetric Higgs Factory}

\author*[a]{Antoine Laudrain}
\author[a]{Ties Behnke}
\author[a]{Carl Mikael Berggren}
\author[a]{Karsten Buesser}
\author[a]{Frank Gaede}
\author[a]{Christophe Grojean}
\author[a]{Benno List}
\author[a]{Jenny List}
\author[a]{Jürgen Reuter}
\author[a]{Christian Schwanenberger}

\affiliation[a]{
    Deutsches Elektronen-Synchrotron DESY, \\
    Notkestrasse 85, 22607 Hamburg, Germany
}

\emailAdd{antoine.laudrain@desy.de}

\abstract{
    The Hybrid Asymmetric Linear Higgs Factory (HALHF) proposes a shorter and cheaper design for a future Higgs factory.
    It reaches a $\sqrt{s} = \SI{250}{\GeV}$ using a \SI{500}{\GeV} electron beam accelerated by an electron-driven plasma wake-field, and a conventionally-accelerated \SI{31}{\GeV} positron beam.
    Assuming plasma acceleration R\&D challenges are solved in a timely manner, the asymmetry of the collisions brings additional challenges regarding the detector and the physics analyses, from forward boosted topologies and beam backgrounds.
    This contribution will detail the impact of beam parameters on beam-induced backgrounds, and provide a first look at what modification compared to e.g. the ILD can improve the physics performance at such a facility.
    The studies will be benchmarked against some flagship Higgs Factory analyses for comparison.
}

\FullConference{42nd International Conference on High Energy Physics (ICHEP2024)\\
18-24 July 2024\\
Prague, Czech Republic\\}

\begin{document}
\maketitle

\section{Introduction}

All major projects of future colliders, circular and linear, share one issue: their high cost.
The most straightforward way of reducing the cost is to decrease the size of the facility.
However, doing so without losing on the beam energy in a linear accelerator requires higher acceleration gradients.
A promising solution to this issue comes from plasma wake-field acceleration (PWFA), which has made significant progress over the past years and is expected to become available for large colliders in the coming ten to fifteen years.

\begin{figure}[htbp]
    \centering
    \includegraphics[width=0.97\textwidth]{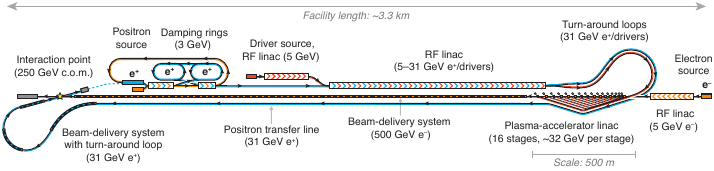}
    \caption{Baseline layout of the HALHF facility~\cite{Foster:2023bmq}.}
    \label{fig:HALHF_layout}
\end{figure}

The Hybrid Asymmetric Linear Higgs Factory (HALHF~\cite{Foster:2023bmq}) proposes a shorter, ILC-style linear collider~\cite{Adolphsen:2013kya} by replacing ILC's radio-frequency acceleration for the electrons by PWFA while keeping the same technology for the positrons acceleration.
The length of the facility can be further reduced by imbalancing the beam energies, putting more energy into the PWFA arm (small length increase) while reducing the energy in the RF arm (large length decrease).
The baseline design for this project is shown in Figure~\ref{fig:HALHF_layout} and features a \SI{250}{GeV} centre-of-mass energy using a \SI{500}{GeV} electron beam colliding with a \SI{31.3}{GeV} positron beam.
The total length of the facility would be in the range of \SIrange{3}{4}{\km} (compared to the \SI{20}{\km} of the ILC) for a cost of around \SI{25}{\percent} of the ILC.

The consequence of the beam energy asymmetry is the introduction of a boost of to the collisions ($\gamma \sim 2.13$ for this set of beam energies).
Designing a detector for such a facility must therefore take into account such boosted topologies.

\section{Beam background constraints}
\label{sec:beam_backgrounds}

The detector design is constrained by the precision required to achieve the physics goals (mainly, but not limited to, Higgs precision physics -- for example $e^+ e^- \to Z(\mu\mu)H$) on the one hand, and by the beam-induced backgrounds on the other hand.
As the physics goals at HALHF match the ones at ILC, the International Large Detector (ILD)~\cite{ILDConceptGroup:2020sfq} was chosen as a starting point.

The beam-induced backgrounds consist in electron-positron pairs created by the electric fields of the crossing beams.
This was extensively studied in~\cite{Laudrain:2023Xp}, where the Guinea-Pig~\cite{Schulte:301970} software was used to simulate the beam background processes.
An updated set of beam parameters was chosen, allowing to place the detector closer to the interaction point and further extended in the forward direction without being saturated or damaged by the beam backgrounds.
The updated set of beam parameters reduced the bunch charge asymmetry to a population of $N_e = \num{3e10}$ electrons and $N_p = \num{1.33e10}$ positrons, while introducing a bunch length asymmetry ($\sigma_{z,e} = \SI{75}{\um}, \sigma_{z,p} = \SI{300}{\um}$).

\begin{figure}[htbp]
    \centering
    \includegraphics[width=0.47\textwidth]{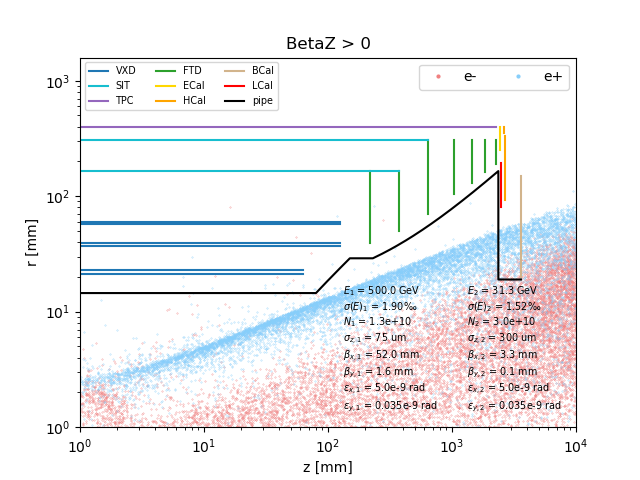}
    \quad
    \includegraphics[width=0.47\textwidth]{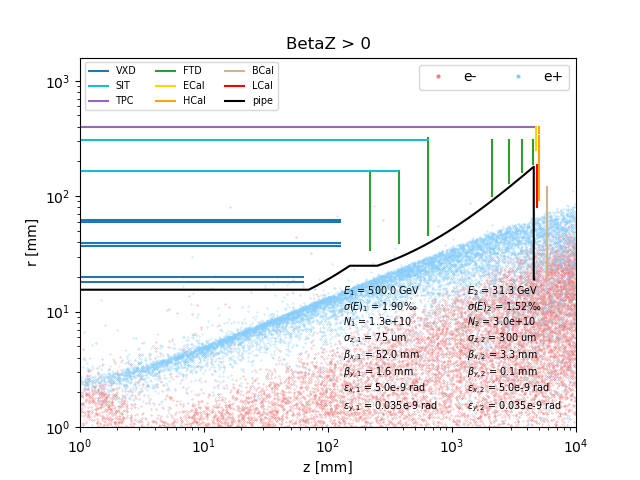}
    \caption{
        Apex of the trajectory of the electron (red) and positron (blue) from the beam backgrounds in asymmetric beams collisions with
        $E_e = \SI{500}{GeV}$, $E_p = \SI{31.3}{GeV}$,
        $N_e = \num{3e10}$, $N_p = \num{1.33e10}$,
        $\sigma_{z,e} = \SI{75}{\um}$, $\sigma_{z,p} = \SI{300}{\um}$.
        in the forward half.
        The beam backgrounds are simulated using Guinea-Pig~\cite{Schulte:301970}.
        On the left, the standard ILD detector (with the exception of using an experimental magnetic field of \SI{5}{\tesla} instead of \SI{3.5}{\tesla}) is superimposed.
        On the right, the "optimised extended ILD" also using an experimental magnetic field of \SI{5}{\tesla} is superimposed.
    }
    \label{fig:pairs_ILD}
\end{figure}

Since the pairs typically have a low transverse momentum, they will spiral in the magnetic field created by the experiment's magnet.
A simple approach to determine whether they hit the detector is to plot the position of the apex of their trajectory in the $x-z$ plane.
Figure~\ref{fig:pairs_ILD} (left) shows such a plot using the above beam parameters set, with a schematic of ILD's innermost parts superimposed.

As also described in~\cite{Laudrain:2023Xp}, studies using ILD's fast simulation (SGV~\cite{Berggren:2012ar}) showed that an extension of a factor of two of the barrel elements length would recover most of the physics performance (based on the Higgs mass measurement benchmark).
Since this "extended-ILD" (e-ILD) layout combined with the above beam parameters set stil gave a few pairs hitting the detector, the detector geometry was improved by
\begin{itemize}
    \item ensuring a \SI{5}{\mm} clearance between the pairs and the beam pipe,
    \item using the doubled TPC size defined in earlier studies (\SI{4700}{\mm} instead of \SI{2350}{\mm}) as well as increased barrel calorimeter subsystems length (not shown in these plots),
    \item extending the vertex detector (VXD) as forward as possible,
    \item rescaling the furthest forward tracking disks (FTD) locations to match the extended TPC length,
    \item and moving the forward calorimeters further downstream.
\end{itemize}
This "improved e-ILD" layout cleared the detector of the bulk of beam backgrounds, even adding a small margin.
Figure~\ref{fig:pairs_ILD} (right) shows the pairs distribution superimposed with this improved e-ILD detector layout.
The focus was put on improving the layout in the forward region only for reasons explained in Section~\ref{sec:HALHF_fullsim}.

\section{Toward a full simulation of a HALHF detector}
\label{sec:HALHF_fullsim}

An other idea to improve the muon momentum resolution is to add another magnetic field in the forward direction to increase the lever arm of the muon track.
SGV can only assume a solenoidal magnetic field, so a full simulation of the improved e-ILD is needed.

\begin{figure}[htbp]
    \centering
    \includegraphics[width=0.49\textwidth]{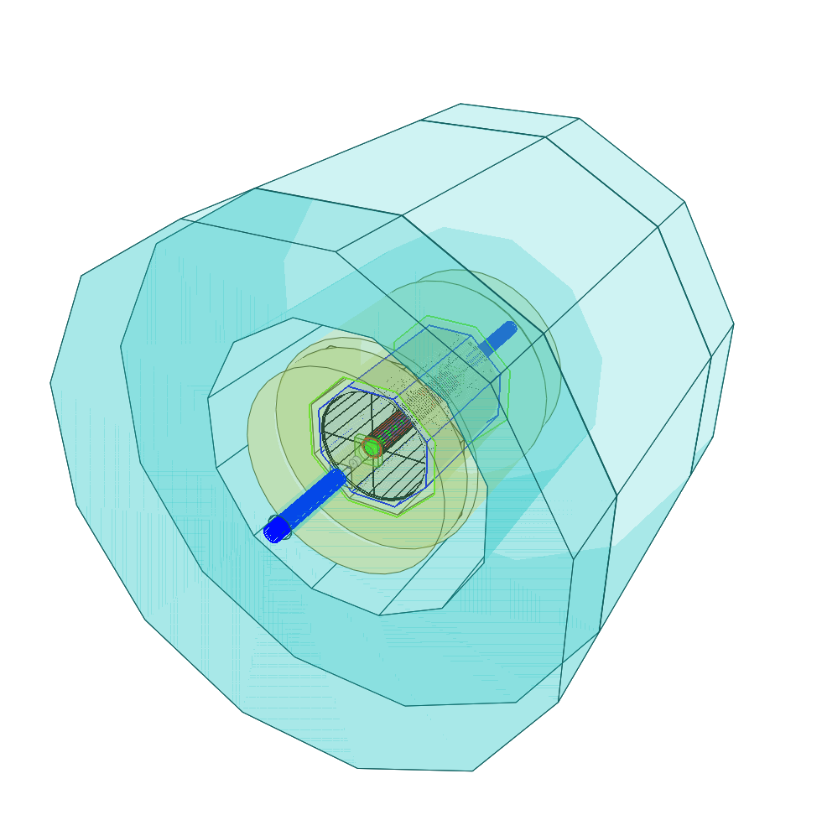}
    \quad
    \includegraphics[width=0.45\textwidth]{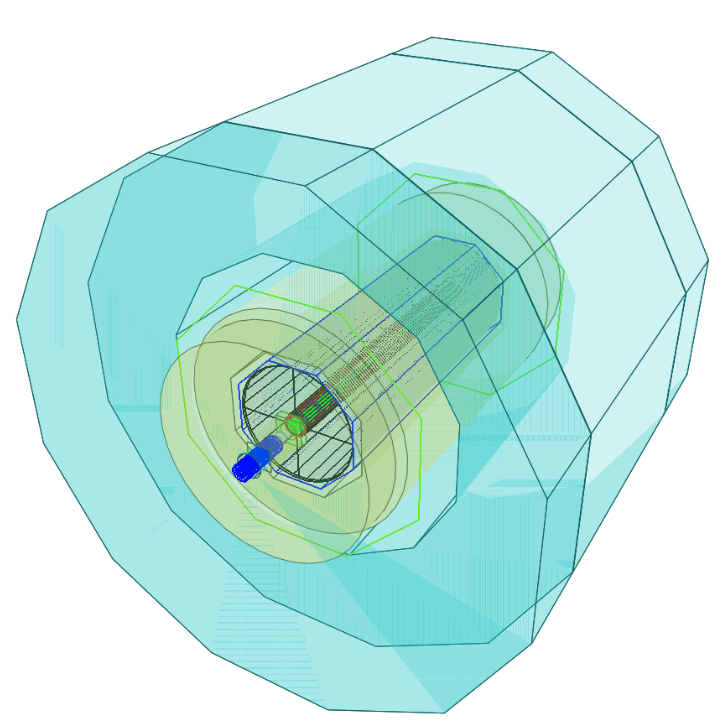}
    \caption{Display of the ILD (left) and the improved e-ILD (right) Geant4 implementation.}
    \label{fig:detector_display}
\end{figure}

\begin{figure}[htbp]
    \centering
    \includegraphics[width=0.49\textwidth]{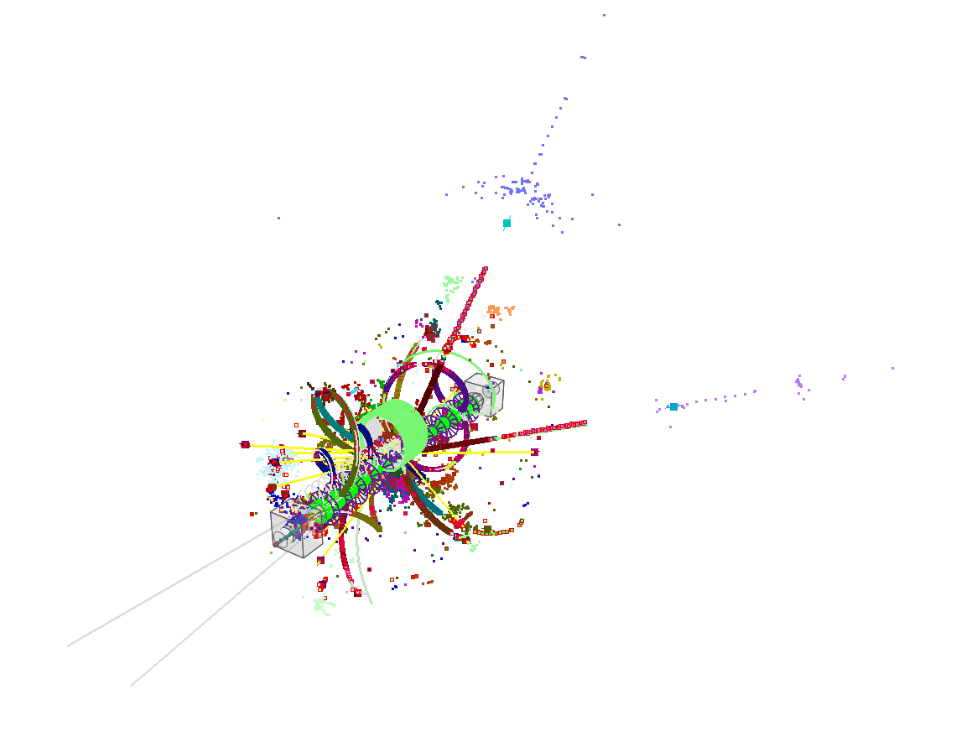}
    \quad
    \includegraphics[width=0.45\textwidth]{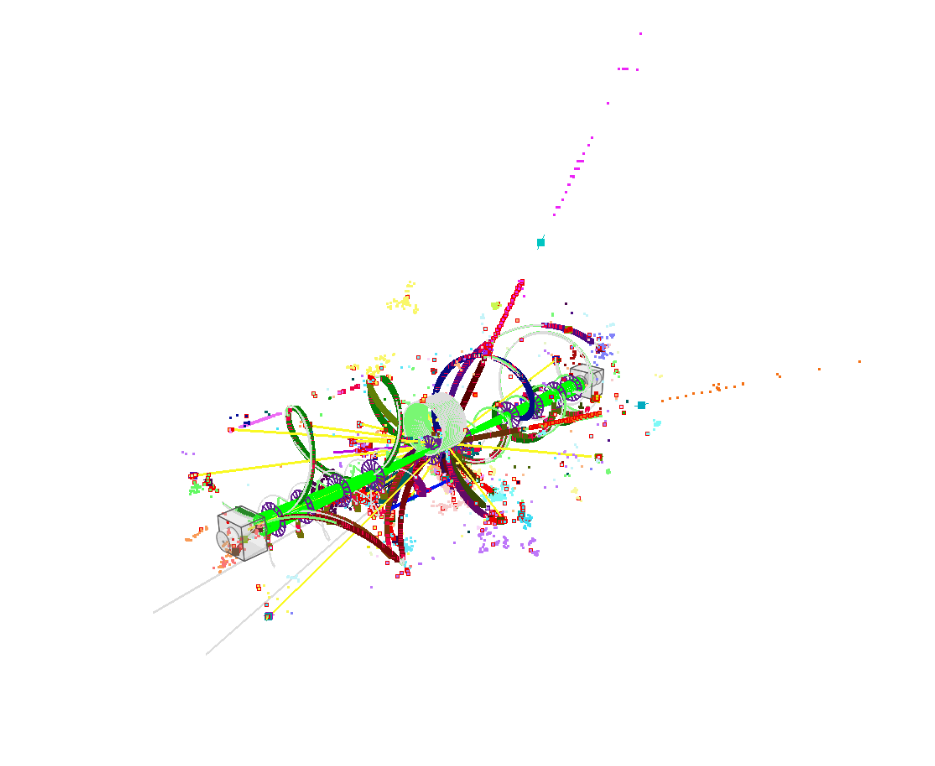}
    \caption{
        Reconstructed $Z(\mu\mu)H$ event in the ILD (left) and in the improved e-ILD (right, same event).
        The beam parameters assumed the ILC configuration, \emph{i.e.} symmetric beams energies (so no boost in the collision) in both cases.
    }
    \label{fig:inner_display}
\end{figure}

The full simulation of the ILD was therefore modified to include the modifications described in Section~\ref{sec:beam_backgrounds}.
However, the original ILD implementation in the DD4HEP framework~\cite{frank_markus_2018_1464634} assumes a symmetric detector.
As a first step, both the forward and backward parts of the ILD layout were modified symmetrically, with the focus only put on the forward section.
A representation of the fully simulated ILD and improved e-ILD detectors in Geant4 is shown in Figure~\ref{fig:detector_display}, nicely showing the extended TPC and barrel calorimeter subsystems.
Since most of the modifications impact the innermost and forward detectors, a close-up in shown in Figure~\ref{fig:inner_display}.
A $Z(\mu\mu)H$ event (with the symmetric ILC beam parameters) is superimposed on the latter figure, clearly showing extended curling tracks in the TPC.

The large simulated dataset for the ILC can be easily used at HALHF by a simple boosting of the particles created during the collisions.
Boosted datasets have been created, and work is ongoing to demonstrate that the full reconstruction works with boosted events in the modified ILD simulation.
This validates the proof-of-concept for exploring further this modified ILD Geant4 simulation by introducing a additional magnetic field in the forward region.

\section{Conclusion}

The HALHF facility could decrease the ecological footprint and cost of a future linear Higgs factory by using plasma wake-field acceleration to accelerate the electrons.
Such a facility introduces asymmetric beam energies, leading to a boost in the collisions which requires an suitable detector.
An initial detector configuration was determined using the ILD fast simulation as a starting point, showing that extending the detector in the forward region recovers most of the performance compared to symmetric collisions.

A detailed study of the beam-induced backgrounds led to a further optimisation of this "extended-ILD" detector layout, which was later implemented in a Geant4 full simulation.
This marks a milestone, as the full simulation allows for modifications of the experimental magnetic field, which is a promising idea to improve the performance of the detector in the forward region.
    Boosting the ILC simulated dataset can provide large samples usable at HALHF, and work is ongoing to show that full reconstruction works with boosted collisions in the modified simulation.

While implementing an asymmetric detector using the modified ILD full simulation is not straightforward, this will be possible in the fast simulation thanks to a recent development in SGV.

\section*{Acknowledgments}

We would like to thank the LCC generator working group for producing the
Monte Carlo samples used in this study.

We thankfully acknowledge the support by the Deutsche Forschungsgemeinschaft (DFG, German Research Foundation) under Germany's Excellence Strategy EXC 2121 "Quantum Universe" 390833306.

This work has benefited from computing services provided by the German National Analysis Facility (NAF)~\cite{Haupt:2010zz}.

We would finally like to thank Martina Mezzolla for her work in this project as a summer student in DESY.

\bibliographystyle{JHEP}
\bibliography{Proceeding_ICHEP2024-Antoine_Laudrain-HALHF}

\end{document}